\documentstyle[multicol,floats,epsfig,aps]{revtex}

\begin{document}
\draft
\title{Oscillatory Exchange Coupling and Positive Magnetoresistance in Epitaxial Oxide Heterostructures}

\author{K.~R.~Nikolaev, A.~Yu.~Dobin, I.~N.~Krivorotov, W.~K.~Cooley,
A.~Bhattacharya, A.~L.~Kobrinskii,
L.~I.~Glazman, R.~M.~Wentzcovitch$^1$,
E.~Dan~Dahlberg, and~A.~M.~Goldman}
\address{School of Physics and Astronomy, University of Minnesota, Minneapolis, MN 55455 \\
$^1$Department of Chemical Engineering and Material Sciences, University of Minnesota, Minneapolis, MN 55455 }

\date{\today}

\maketitle

\begin{abstract}

Oscillations in the exchange coupling between ferromagnetic $La_{2/3}Ba_{1/3}MnO_3$ layers
with paramagnetic $LaNiO_3$ spacer layer thickness has been observed in epitaxial heterostructures
of the two oxides. This behavior is explained within the RKKY model 
employing an {\it ab initio} calculated band structure of $LaNiO_3$, taking into account
strong electron scattering in the spacer.  Antiferromagnetically coupled 
superlattices  exhibit a positive current-in-plane magnetoresistance.
\end{abstract}


\begin{multicols}{2}
Since the discovery of giant magnetoresistance~\cite{baibich88}
and oscillatory interlayer coupling~\cite{parkin90},
metallic magnetic multilayers have been the subject of intensive research~\cite{ibmrev98}.
A physical picture of the coupling is provided in terms of quantum
interference due to confinement of electrons in the nonmagnetic
spacers~\cite{stiles93,bruno95}.  Ruderman-Kittel-Kasuya-Yosida (RKKY)
theory~\cite{bruno92}, successfully used to describe the effect,
appears to be a limiting case of a more general
approach~\cite{bruno95}.  The prediction that oscillations periods are
determined by the extremal spanning vectors of the Fermi surface of
the nonmagnetic spacer is now supported by a wealth of experimental
data~\cite{stiles99}.  The phase and magnitude of the coupling
oscillations apparently are sensitive to the interface matching of the
electron bands for a particular magnetic configuration of the
layers~\cite{ibmrev98,stiles93,bruno95}.

While these effects have been studied in many simple metal and alloy systems, 
little progress has been achieved in investigating them in multilayers 
consisting entirely of compounds~\cite{orozco99}.  
In this Letter we report the first observation of 
oscillatory coupling in heterostructures fabricated entirely of oxides, 
thus extending the field to a novel 
class of materials. The ferromagnetic layer material, barium-doped lanthanum manganese oxide 
$La_{2/3}Ba_{1/3}MnO_3$ belongs to the family of metallic manganese oxides 
that exhibit very large (colossal) magnetoresistance~\cite{ramirez97}.
Double exchange ferromagnetism~\cite{degennes60} predicts
half-metallicity in these compounds, which 
has been justified by {\it ab initio} band structure calculations~\cite{singh98} as well as confirmed 
(to a certain degree) experimentally~\cite{soulen98}. 
As a spacer layer material, we have employed $LaNiO_3$, which is lattice-matched to 
$La_{2/3}Ba_{1/3}MnO_3$, and is the only rare earth nickelate  that is a paramagnetic metal~\cite{medarde97}. 
Its susceptibility is strongly enhanced by electron-electron exchange interactions~\cite{sreedhar92}.  In thin 
film form, it is a better conductor (resistivity of 
$50-100 \ \mu\Omega \cdot cm$) than the manganites (resistivity of 
$300-500 \ \mu\Omega \cdot cm$).  Although advances in oxide film 
growth permit the fabrication of atomically defined layered structures with nanometer scale 
periodicity~\cite{izumi99}, the very strict control over stoichiometry and deposition conditions 
required to produce multilayers remains a significant experimental challenge.

In our previous paper~\cite{nikolaev99} we presented the evidence for antiferromagnetic (AFM)
interlayer coupling in this system. Here we demonstrate that the initially AFM coupling
becomes ferromagnetic (FM) at larger spacer thicknesses and explain this behavior
within the RKKY model.

Superlattices were grown on heated $SrTiO_3(001)$ substrates by
block-by-block molecular beam epitaxy (MBE)~\cite{locquet94} in the
presence of an ozone flux, using previously
described procedure~\cite{achutharaman92}. The RHEED patterns observed
during the block-by-block deposition, indicated smooth surfaces at the completion of
each monolayer throughout the entire process of growth.
Characterization by high resolution X-ray diffraction showed that the
films were monocrystalline with $c$-axis oriented perpendicular to the substrate.
Well-defined satellite peaks indicated the sharpness of
the interfaces and good superlattice periodicity~\cite{nikolaev99}.
Cross-sectional transmission electron microscopy demonstrated the
continuity of the layers over macroscopic scales, and notably
defect-free structures on atomic length scales.

For a single very thin manganite film encapsulated between two nickelate layers, 
hysteresis loops with the field applied in various directions in the (001) plane of the film 
can be described by a biaxial magnetocrystalline anisotropy. 
This demonstrates that domain-wall pinning is negligible and justifies  a model of coherent
rotation of the magnetization in the analysis of 
hysteresis.

\begin{figure}
\epsfig{file=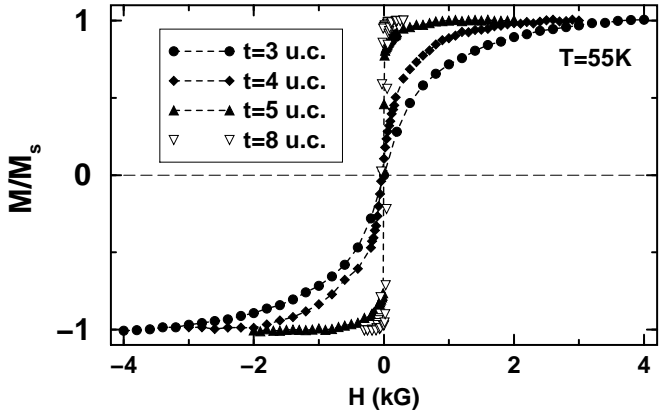,width=3.4in}
\caption{Hysteresis loops for a series of 
($La_{2/3}Ba_{1/3}MnO_3$ 12 u.c. / $LaNiO_3$ $t$)$_{10}$  superlattices
obtained using a superconducting susceptometer.}
\label{fig_hyst}
\end{figure}

Figure \ref{fig_hyst} shows typical M-H curves for 
a series of 
($La_{2/3}Ba_{1/3}MnO_3$  / $LaNiO_3$)$_{10}$ (001)  superlattices.  The thickness
of each manganite layer was fixed at 12 unit cells (u.c.), while that of nickelate varied
from 3 u.c. to 10 u.c.~\cite{unitcell}. For protection, 
the structures were encapsulated between two $50 \AA$-thick  nickelate layers.
\end{multicols}
\twocolumn
\noindent
The magnetic field was directed along the [100] direction, coinciding
with the easy axis 
of the manganite films.
Samples with spacer layers three- and four-unit-cell thick, showed zero remanence and high saturation 
fields typical of AFM coupling. The spontaneous magnetizations 
of these samples at low temperatures were less than one percent of saturation, 
illustrating that all the periods have the same thickness and composition, and that there were 
no pinholes bridging between layers.
However, the shape of the magnetization curves suggests a substantial biquadratic 
contribution to the interlayer coupling~\cite{demokritov98}. For thicker spacer structures, 
hysteresis loops had
high remanence and the saturation fields were much lower.  The coupling constants were
estimated by fitting hysteresis curves with those calculated from a 
model with coherent rotations~\cite{fitting}. 

\begin{figure}[!t]
\epsfig{file=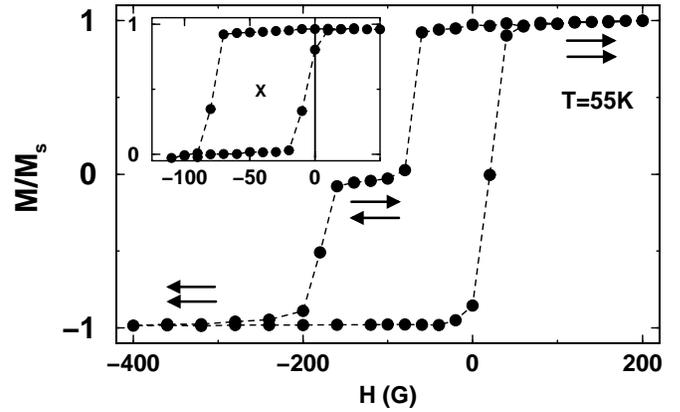,width=3.4in}
\caption{Hysteresis loop for a ''spin-valve''-like structure with eight-unit-cell thick
$LaNiO_3$ spacer obtained using a superconducting susceptometer. Arrows indicate magnetization directions. 
Inset: minor hysteresis loop.}
\label{fig_spinvalve}
\end{figure}

To estimate the coupling constants of small magnitude anticipated for
structures with thicker spacer layers, a series of samples with varied
spacer layer thickness $t$ were grown in a
''spin-valve''~\cite{dieny91} four-layer geometry:
$La_{2/3}Ba_{1/3}MnO_3$ 12 u.c. / $LaNiO_3$ $t$ /
$La_{2/3}Ba_{1/3}MnO_3$ 12 u.c. / $La_{1/3}Ca_{2/3}MnO_3$.  In this
configuration the magnetization of one manganite layer is free to
rotate while the other one is pinned by the 150\AA-thick epitaxial
layer of the antiferromagnet,
$La_{1/3}Ca_{2/3}MnO_3$~\cite{nikolaev00}.  Measurements were
performed on field-cooled films with a magnetic field applied along
the [100] direction.  As an example, a hysteresis loop of a structure
with an eight-unit-cell thick spacer is shown in
Fig.~\ref{fig_spinvalve}.  The inset shows the hysteresis loop
corresponding to an unbiased FM layer ("minor" loop). The magnetic
coupling between the manganite layers leads to a change of an external
field at which the magnetization of the unbiased layer flips. This
results in a shift of the minor loop along the field axis, the sign
and magnitude of the shift being a quantitative measure of the
interlayer coupling. No such displacement was detected for samples
with thicker spacers, demonstrating that the magnetic layers are
decoupled. For nickelate spacers six-and seven-unit-cell thick single
hysteresis loops were observed with the magnetizations of both layers
reversing their orientations simultaneously. This is indicative of a
strong FM interaction. A lower limit on the strength of the coupling
can be set by the difference of the coercive fields of the biased and
free FM layers. In agreement with the superlattice data, the
hysteresis loop of the structure with a thin (four-unit-cell thick)
spacer layer reflected the presence of strong bilinear AFM and
biquadratic couplings.

\begin{figure}[!b]
\epsfig{file=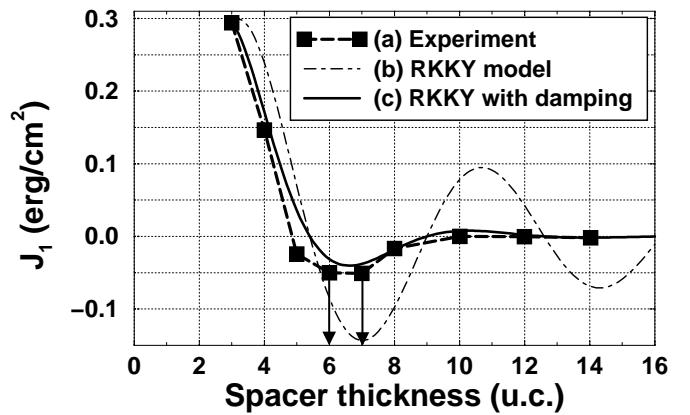,width=3.4in}
\caption{Spacer thickness variation of experimental (curve $a$) and calculated (curves $b$ and $c$) 
coupling strength.
The theoretical curve is $J_1(t) = A \cos(2\pi t /T_{LP}) / t \cdot \exp(-t/\lambda)$
with long period $T_{LP}=7.2$~u.c. and damping  length $\lambda=3$~u.c. as well
as without damping ($\lambda=\infty$). The amplitude 
$A$ is chosen to match the experimental value at $t=3$~u.c.}
\label{fig_depend}
\end{figure}

The combined data from both sets of measurements 
were used to obtain the variation of bilinear coupling strength $J_1$ with  
spacer layer thickness (curve $a$ in Fig.~\ref{fig_depend})~\cite{wedge_disc}. 
As one can see, the initially AFM coupling changes sign 
before vanishing. The coupling strength is comparable to that found in many conventional systems.

\begin{figure}[!t]
\epsfig{file=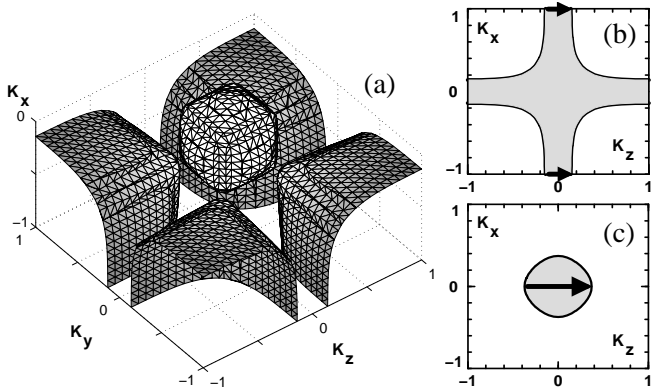,width=3.4in}
\caption{ {\bf (a)}  Fermi surface of $LaNiO_3$ consists of a small $\Gamma$-point
electron sheet and a huge $R$-point hole pocket (half of it is displayed). 
\ \  {\bf (b)} $k_y=\pi/a_0$,  {\bf (c)} $k_y=0$  critical cross-sections of the Fermi surface.
Arrows indicate extremal spanning wave vectors. The wave vectors are in units of $\pi/a_0$.}
\label{fig_fs}
\end{figure}

A conventional RKKY model with additional 
damping caused by strong electron scattering in the nonmagnetic layer has been used to 
describe the observed variation of the coupling.  The topography 
of the Fermi surface of the nickelate spacer is vital to the description of the exchange 
coupling. The band structure and density of states of $LaNiO_3$ have been computed using the 
pseudo-potential plane wave approach~\cite{calc_method}, and are in good agreement with previous 
calculations~\cite{solovyev96}.  The Fermi surface of $LaNiO_3$ is depicted in Fig.~\ref{fig_fs}$a$.  
For the [001] 
direction, there are two apparent extremal spanning vectors associated with long 
$T_{LP}=7.2$~u.c. (Fig.~\ref{fig_fs}$b$) and short $T_{SP}=2.7$~u.c. (Fig.~\ref{fig_fs}$c$) 
oscillation periods.  
Due to a much larger radius of curvature of the Fermi surface $K_F$ at the long period extremal 
spanning vector, its contribution is dominant.  In addition, short periods are expected 
to be suppressed by interface roughness.  Because of the nested-like feature of the Fermi 
surface, the asymptotic RKKY formula is not applicable for thin ($t < K_Fa_0^2 \sim 40$~u.c.) 
spacers.  Considerations similar to ones used in the case of complete planar nesting~\cite{bruno92}
significantly modify the spacer thickness dependence of the coupling strength which is in 
this case given by: $J(t) \propto - \cos (2\pi t /T_{LP})/t$, where $t$ is the spacer thickness.
The comparison with experimental data (curves $a$ and $b$ in Fig.~\ref{fig_depend}) 
reveals good agreement in both period and phase. However, the experimentally observed 
decay rate of the coupling strength is faster than predicted.  We argue that strong 
electron scattering in $LaNiO_3$ may account for the oscillation damping 
of the form of $\exp(-t/\lambda)$, where $\lambda$ is the scattering length 
of the "coupling" electrons.  An estimate for  $\lambda=3$~u.c.~\cite{lambda_calc} 
obtained from the measured resistivity describes the 
observed damping reasonably well (see curve $c$ in Fig.~\ref{fig_depend})~\cite{eeinteract}.  

\begin{figure}[!t]
\epsfig{file=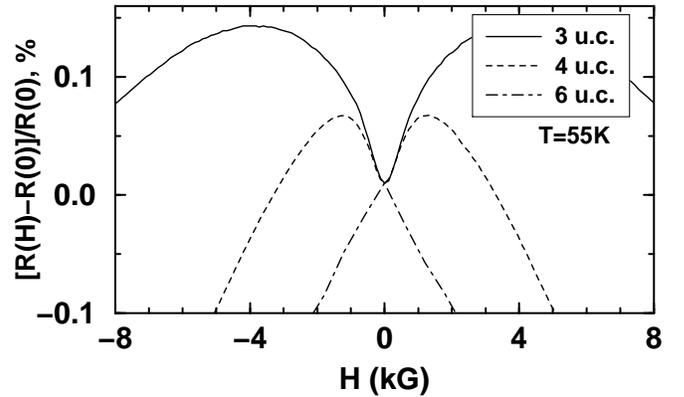,width=3.4in}
\caption{Magnetoresistance vs in-plane magnetic 
field dependence for the ($La_{2/3}Ba_{1/3}MnO_3$ 12 u.c. / $LaNiO_3$ $t$)$_{10}$  superlattices. 
Actual magnetoresistance is higher since part of the current is 
shunted by $LaNiO_3$ capping layers.}
\label{fig_mr}
\end{figure}

Current-in-plane magnetoresistance (MR) measurements were made using a standard 
four-probe DC technique.  Results are shown in Fig.~\ref{fig_mr} for three superlattices with spacer 
layer thicknesses of three, four, and six unit cells, respectively.  The observed decrease in resistance at 
large fields originates from the negative MR of the manganite 
layers.  Unlike the case of bulk material and thicker films, substantial 
negative MR is observed in single ultrathin manganite films well below the Curie 
temperature~\cite{izumi98}. The low-field positive MR is present only in structures 
that exhibit AFM interlayer coupling.  
The resistance has a minimum for the AFM configuration,
whereas a maximum is seen when the layers become FM aligned.  
This provides compelling evidence that the phenomenon is associated with the 
configuration of the magnetizations of adjacent layers.  Measurements performed with a 
field applied at various angles to the direction of current yielded the same result,
thus demonstrating the effect is not caused by in-plane MR anisotropy.
The positive MR we found is in contrast with the negative magnetoresistance
conventionally observed in transition metal 
multilayers.  In latter structures the effect of the positive sign is only observed in 
systems with spin asymmetries in the electron scattering in successive FM layers~\cite{george94}.

We speculate that this behavior may be caused by the intrinsic negative MR of the manganite
layers. In this model, an indirect exchange interaction affects the transport in the manganites~\cite{krivorotov00}. 
If one views the superlattice as a parallel
resistor network, the low-field positive MR can indeed be achieved if the coupling is
AFM with a biquadratic contribution. Alternatively, the effect may originate from
the variation of the electronic structure of the system due to the change in the magnetic configuration.

In conclusion, heterostructures consisting of layers
of a colossal magnetoresistance material separated by a nonmagnetic
metallic oxide spacers have been successfully grown by MBE method. The high
quality of the heterostructures allowed us for the first time to study the magnetic
interlayer coupling, and demonstrate the oscillatory dependence of the
coupling strength on the spacer thickness.
Exchange coupling in this exotic system has been described within
the RKKY model.  The origin of the positive magnetoresistance with a
cusp-like feature at small magnetic fields remains unclear at this
time.

This work was supported by the NSF through the {MRSEC} program, Grant
No.~NSF/DMR-9809364, by the NSF under Grants No.~DMR-9731756 and DMR-9812340,
and by the ONR under Grant No.~N00014-98-1-0098.


\begin{references}
\bibitem{unitcell}
In-plane lattice constants of $3.905 \AA$ are set by the substrate.
Out-of-plane lattice constants for manganite and nickelate films
are $3.94 \AA$ and $3.81 \AA$, respectively.
\bibitem{fitting}
The total energy of the system is given by:
$
      E=\sum_j J_0/2 \cos (\phi_j-\phi_{j+1})+ J_1/2 \cos 2(\phi_j-\phi_{j+1})
      -HM \cos(\phi_j)+ K/2 cos(4\phi_j)
$,
where the first two terms are the  bilinear and biquadratic couplings, respectively, 
and the last two are the 
Zeeman and magnetocrystalline four-fold anisotropy terms.
The $\phi_j$ are in-plane orientation angles of the magnetizations.  
Numerically minimizing this energy leads to a model curve for fitting to the data.
However, coupling constants cannot be 
determined unambiguously from a fit when they are small 
compared to the magnetocrystalline anisotropy, or when the biquadratic coupling constant 
is smaller than the ferromagnetic bilinear one.
\bibitem{wedge_disc}
Of course fabrication of a wedged sample would be desirable, but the need to 
maintain a uniform substrate temperature and ozone flux during growth makes this 
approach, so successful with metals, not possible.
\bibitem{calc_method}  
We used for these calculations norm-conserving  pseudopotentials~\cite{troullier91}
with partial core correction and the generalized gradient 
approximation~\cite{perdew96} for exchange-correlation potential. To test  the validity of
our approach we have optimized lattice parameters for bulk $LaNiO_3$  (rhombohedral  structure)
and found them to be in a good agreement with experimental ones~\cite{garcia92}. In the present 
experiment the compound is tetragonally distorted (fully strained film). The out-of-plane
lattice  parameter obtained by optimization in such a constrained situation is very close to the one
obtained from  our X-ray diffraction measurements on fully strained nickelate single film grown on
$SrTiO_3$ (001) substrate.
\bibitem{lambda_calc}
We obtained the scattering time $\tau$ (assumed to be  isotropic) from
the resistivity of a thick film.
The coupling decay length is then given by $\lambda=v_{z} \cdot \tau$,
where $v_{z}$ is the velocity at the extremal wave vector.
\bibitem{eeinteract}
The strong electron-electron correlations in the spacer 
layer can give rise to an additional ferromagnetic coupling.  This has been studied 
both experimentally~\cite{gelinski90} and theoretically~\cite{takahashi97} 
in $Pd$-based intermetallic layered structures in 
which it was found to be very strong, even completely eliminating the regimes of 
antiferromagnetic coupling.  Like $Pd$, $LaNiO_3$ exhibits enhancement of both its 
electronic magnetic susceptibility and specific heat~\cite{sreedhar92}. 
 An analysis of this effect in our 
structures is complicated, in part due to the necessity of calculating the interaction in 
the crossover from the ballistic~\cite{takahashi97} to the diffusive~\cite{altshuler83}
 regimes.  There is no 
independent evidence to suggest that strong electron-electron interaction significantly 
modifies the interlayer coupling in our structures. The RKKY interaction alone produces an 
acceptable description of the coupling variation within the accuracy of the data.
\end{references}
\end{document}